\documentclass[prb,aps,twocolumn,showpacs]{revtex4}
\usepackage{amsmath,amssymb,bm}
\usepackage{graphicx}

\DeclareMathOperator{\im}{Im}

\begin{document}

\author{Anna Sitek}
 \email{anna.sitek@pwr.wroc.pl}
\author{Pawe{\l} Machnikowski}
 \email{pawel.machnikowski@pwr.wroc.pl}
 \affiliation{Institute of Physics, Wroc{\l}aw University of
Technology, 50-370 Wroc{\l}aw, Poland}

\title{Theory of nonlinear optical response of ensembles of double quantum dots}

\begin{abstract}
We study theoretically the time-resolved four-wave mixing (FWM) response of an
ensemble of pairs of quantum dots undergoing radiative recombination. 
At short (picosecond) delay times,
the response signal shows beats that may be dominated by the
subensemble of resonant pairs, which gives access to the information of
the inter-dot coupling. At longer delay times, the decay of the FWM
signal is governed by two rates which result from the collective
interaction between the two dots and the radiation modes. The two rates
correspond to the subradiant and superradiant components in the
radiative decay.
Coupling between the dots enhances the collective effects and makes
them observable even when the average energy mismatch between the dots
is relatively large.
\end{abstract}


\maketitle

\section{Introduction}

Double quantum dots (DQDs) are pairs of quantum dots (QDs) placed at a
small distance from each other and coupled either by long-range dipole
forces \cite{lovett03b,danckwerts06} or by carrier tunneling through
the inter-dot barrier \cite{bryant93,schliwa01,szafran01} (in the
latter case, the system is referred to as a quantum dot
molecule). Such systems 
attract much attention both in the
theoretical and experimental research. This still growing interest is
driven to a large extent by the technological promise these structures 
show for nanoelectronics and quantum information processing
applications. A system built of two QDs can be viewed as a first step
towards scalable semiconductor-based quantum devices. Coupling
between the dots is essential here, as it allows one to induce
conditional dynamics in the system and thus to realize the basic elements
of quantum computing
\cite{biolatti00}. 
Another interesting application of coupled QDs is generation of
entangled photons \cite{gywat02}.
It is therefore not surprising that much
experimental work has been devoted to proving the existence of
coupling in DQD systems
\cite{ortner05,bayer01,ortner03,ortner05b,krenner05b}. 

In this paper, we study the four-wave mixing (FWM) optical response of an
inhomogeneous ensemble of self-assembled double quantum dots. The
systems under consideration are QD pairs formed spontaneously by
strain-induced nucleation, as is typical, e.g., in the InAs/GaAs system
\cite{xie95,solomon96} (although, in fact, the exact arrangement of
the QDs is not essential in our discussion). 
The FWM spectroscopy is an optical technique
commonly used for extracting
information on lifetimes and homogeneous dephasing from inhomogeneous
ensembles \cite{borri99,borri01,vagov04}. It has been also applied to DQD
ensembles \cite{borri03}, showing features that are clearly distinct
from those observed in ensembles of individual QDs, like two-component
decay and modified initial dephasing. As the physical properties of
DQDs are much more complex than those of individual QDs it is not
always possible to identify the mechanism responsible for these 
differences. In particular, it is not clear to which extent they may
result from simple optical interference or
quantum-optical mechanisms of collective radiative decay
(sub- and superradiance), as opposed, e.g., to DQD-specific phonon-related
dephasing channels, like dissipative exciton transfer
\cite{govorov03,rozbicki08a}. 

Superradiance has been thoroughly studied for strongly excited atomic
samples \cite{gross82,skribanovitz73}, where it is manifested as an
outburst of radiation due to constructive interference of
quantum-mechanical amplitudes for radiative transition in a massively
correlated state of atoms interacting with a common radiation
reservoir. Somewhat less spectacular form of this effect was also
observed in QD ensembles \cite{scheibner07} where the radiative
recombination was shown to increase for a sufficiently large number of
dots. However, the essential features of the collective emission can
be observed already in a two-emitter system: states with a delocalized
excitation decay slower or faster, depending on the relative phase
between the two components in the superposition (the first or the
second emitter excited) \cite{agarwal74,sitek07a}.

The goal of the present work is to identify the purely optical effects
that may appear in the nonlinear optical response of an ensemble of
DQDs. It has been pointed out that the dynamics of double dots coupled
to a radiative reservoir is very reach, both in the open space
\cite{sitek07a} and in a cavity \cite{hughes05}.
Here, we will show that the exponential decay
of optical coherence, characteristic of a
single QD, is replaced by a non-exponential, two-rate decay which may
be related to sub- 
and superradiant components in the system evolution. As in the
previously studied case of a single pair of nearly identical QDs
\cite{sitek07a}, the decay is strongly affected by the coupling
between the dots. We show that the collective features in the radiative decay
persist even for relatively large values of the energy mismatch
between the dots (many orders of magnitude larger that the emission
line width) as long as the coupling is of comparable
magnitude. Another feature that emerges from our model is a damped
oscillatory behavior of the FWM signal at short
delays (sub-picosecond and picosecond time scales), 
which results from the optical interference of the signals
emitted by various DQDs in the ensemble, dephased due to inhomogeneity
of DQD parameters.
As we will show, these beats may be dominated by a relatively
long-living contribution from 
the minority subensemble of resonant DQDs (formed by dots with identical
transition energies). In this way, the nonlinear response contains
information on the coupling between the dots, irrespective of the
energy mismatch between them.

The paper is organized as follows: In sec. \ref{sec:system} we present
the system under consideration and define its model. Next, in
sec. \ref{sec:evol} we discuss the evolution of the system under
two-pulse optical driving and derive the
equations for the third order optical polarization. The results are
discussed in sec. \ref{sec:results}. Sec. \ref{sec:concl} contains
concluding remarks. 
	
\section{The system}
\label{sec:system}

We study an ensemble of DQDs, each consisting of two QDs. 
The QDs in each pair are placed at a distance much smaller 
than the relevant photon wavelength so that spacial dependence of the 
electromagnetic (EM) field may be neglected (the Dicke limit). We
assume that the polarization of the laser beam is chosen such that
only one of the fundamental excitonic transitions in each dot is
allowed. 
Each DQD can be then modelled as a
four-level system with the basis states
$|0\rangle\equiv|00\rangle$, $|1\rangle\equiv|01\rangle$, 
$|2\rangle\equiv|10\rangle$, $|3\rangle\equiv|11\rangle$, corresponding to the
ground state (empty dots), an exciton in the second and first QD, 
and excitons in both QDs, respectively.
A single DQD is
composed of two QDs with energies $E_{1,2}=E\pm\Delta$.
We will describe its evolution in a frame rotating with the frequency $E/\hbar$.
Then, in the rotating wave approximation, the Hamiltonian for a single
DQD is 
\begin{equation*}
	H=H_{\mathrm{X}}+H_{\mathrm{L}}+H_{\mathrm{rad}}+H_{\mathrm{phot}}.
\end{equation*}

The first component describes  the excitons
\begin{eqnarray}
\label{ham}
\lefteqn{H_{X}=}\nonumber\\
&&\Delta\left(\hat{n}_{1}- \hat{n}_{2}\right)
+V\left( \sigma^{\left(1\right)}_{-}\sigma^{\left(2\right)}_{+} 
+ \sigma^{\left(1\right)}_{+}\sigma^{\left(2\right)}_{-} \right)
+V_{B}\hat{n}_{1}\hat{n}_{2},	
\end{eqnarray}
where 
$\sigma^{\left(j\right)}_{\pm}$ are creation and annihilation 
operators of an exciton in the j-th QD, 
$\hat{n}_{j}=\sigma_{+}^{(j)}\sigma_{-}^{(j)}$ is the occupation of
the dot $j$, $V$ is the coupling 
between the dots (e.g., tunneling or F{\"o}rster), and $V_{B}$ is 
a biexciton shift due to a static dipole interaction
\cite{danckwerts06}. 

The second term in the Hamiltonian accounts for the interaction with the 
laser field, which is treated classically,
\begin{equation}
\label{ham1}
H_{\mathrm{L}}=\frac{1}{2}\sum_{l}f_{l}(t)\left[ 
e^{-i(\phi_{l}+Et_{l}/\hbar)}
\left(\sigma^{\left(1\right)}_{-}+\sigma^{\left(2\right)}_{-} \right) 
+\mathrm{H.c.} \right],
\end{equation}
where $f_{l}$, $\phi_{l}$, and $t_{l}$ are the amplitude envelopes,
phases, and arrival times of the laser pulses, respectively. 

The third term accounts for the interaction with the quantized EM
field (radiation reservoir) in the dipole approximation.  
\begin{equation*}
H_{\mathrm{rad}}
=\left(\sigma^{\left(1\right)}_{-}+\sigma^{\left(2\right)}_{-}\right)
\sum_{\bm{k},\lambda}g_{\bm{k}\lambda}
e^{i\left(\omega_{\bm{k}}-E/\hbar\right)t}b^{\dagger}_{\bm{k},\lambda}+\mathrm{H.c.},
\end{equation*}
where 
\begin{equation*}
g_{\bm{k}\lambda}=i\bm{d}\cdot\hat{e}_{\lambda}
\left(\bm{k}\right)\sqrt{
\frac{\hbar\omega_{\bm{k}}}{2\epsilon_{0}\epsilon_{r}v}},
\end{equation*}
$\bm{k}$ is a photon wave vector, $\omega_{\bm{k}}$ is the
corresponding frequency, $\lambda$ denotes polarizations,
$b_{\bm{k},\lambda},b_{\bm{k},\lambda}^{\dag}$ are photon creation and
annihilation operators,
$\bm{d}$ is the interband dipole moment (for simplicity equal for 
all QDs), $\hat{e}_{\lambda}\left(\bm{k}\right)$ is a unit polarization 
vector, $\epsilon_{0}$ is the vacuum permittivity,
$\epsilon_{\mathrm{r}}$ is the dielectric constant of the semiconductor,
and $v$ is the normalization volume for the EM modes. 

Finally,
\begin{eqnarray}
	H_{\mathrm{phot}}=\sum_{\bm{k},\lambda}\hbar\omega_{{\bm{k}}}
b_{\bm{k},\lambda}^{\dag}b_{\bm{k},\lambda}
\end{eqnarray}
 is the Hamiltonian of the
radiation reservoir.

To describe the ensemble of DQDs,
we assume a Gaussian distribution function
for the energies of the two dots,
\begin{eqnarray}\label{g}
g(E_{1},E_{2})& = &
\frac{1}{2\pi\sigma^{2}\sqrt{1-\rho^{2}}}\nonumber \\
&& \times e^{-\frac{\left(E_{1}-\bar{E}_{1}\right)^{2}
-2\rho\left(E_{1}-\bar{E}_{1}\right)\left(E_{2}-\bar{E}_{2}\right)
+\left(E_{2}-\bar{E}_{2}\right)^{2}}{2\left(1-\rho^{2}\right)\sigma^{2}}}
\end{eqnarray}
with the mean transition energies $\bar{E}_{1}$, $\bar{E}_{2}$,
identical energy variances $\sigma^{2}$ for both QDs, and a correlation
coefficient $\rho$. Note that this distribution corresponds to an 
uncorrelated Gaussian distribution of the parameters $E$ and 
$\Delta$, $g\left(E,\Delta\right)=g_{E}
\left(E\right)g_{\Delta}\left(\Delta\right)$,
where
\begin{equation}
g_{A}\left(A\right)=
\frac{1}{\sqrt{2\pi}\sigma_{A}}
\exp\left[-\frac{\left(A-\overline{A}\right)^{2}}{2\sigma_{A}^{2}}\right],\quad
A=E,\Delta,
\label{gA}
\end{equation}
with the mean values $\bar{E}=(\bar{E}_{1}+\bar{E}_{2})/2$ and 
$\bar{\Delta}=(\bar{E}_{1}-\bar{E}_{2})/2$ and variances 
$\sigma_{E}^{2}=\sigma^{2}(1+\rho)/2$ and 
$\sigma_{\Delta}^{2}=\sigma^{2}(1-\rho)/2$
(correlation between the QD energies $E_{1},E_{2}$ means less variance
of their difference).

\section{The system evolution and FWM response}
\label{sec:evol}

A FWM experiment, which we want to model, consists in exciting an 
ensemble of DQDs with two ultrashort laser pulses, arriving at 
$t_{1}=-\tau$ and $t_{2}=0$. The first step of the calculation is to
find the optical polarization of a single DQD after the second
pulse, which is  proportional to 
\begin{equation*}
P(t)=\rho_{10}(t)+\rho_{20}(t)+\rho_{31}(t)+\rho_{32}(t)
+\mathrm{c.c.},
\end{equation*}
where $t>0$, $\rho(t)$ is the density matrix of a DQD structure, and 
$\rho_{kl}=\langle k|\rho|l\rangle$. The first two terms are exciton
coherences (polarizations) while the other two are referred to as
biexciton polarizations. In order
to extract the FWM polarization we pick out only the terms containing 
the phase factor $e^{i(2\phi_{2}-\phi_{1})}$, which mimics the
frequency shifting and lock-in detection technique used in the
actual experiment \cite{mecozzi96,borri99}.
In the second step, the total optical response from the sample is
obtained by summing up 
the contributions from individual DQDs with the weight factor
$g(E,\Delta)$,
\begin{equation*}
P_{\mathrm{FWM}}(t)=\int d\Delta dEg(E,\Delta)P(t).
\end{equation*}

\subsection{Single DQD evolution}

The detection of weak signals originating from the DQD ensemble is based
on a heterodyne technique \cite{borri99}: The response
$P_{\mathrm{FWM}}$ is superposed onto a reference pulse
\begin{displaymath}
E_{\mathrm{ref}}(t)=f_{\mathrm{ref}}(t-t_{0})e^{-i\overline{E}(t-t_{0})/\hbar}
+\mathrm{c.c.},
\end{displaymath}
arriving at a time $t_{0}$. We assume a Gaussian envelope
\begin{displaymath}
f_{\mathrm{ref}}(t)=
\frac{1}{\sqrt{2\pi}\tau_{\mathrm{ref}}}
\exp\left[-\frac{1}{2}\left(\frac{t}{\tau_{\mathrm{ref}}}\right)^{2}\right].
\end{displaymath}
The measured signal is proportional to $|F(t_{0},\tau)|$, where
\begin{equation}
\label{F}
F(t_{0},\tau)=e^{i\overline{E}(t_{0}-\tau)/\hbar}\int dt 
P_{\mathrm{FWM}}^{(+)}(t)E_{\mathrm{ref}}^{(-)}(t),
\end{equation}
where $P_{\mathrm{FWM}}^{(+)}$ and $E_{\mathrm{ref}}^{(-)}$ are the
positive frequency part of the FWM signal and the negative frequency
part of the reference pulse, respectively, and the (irrelevant) phase
factor has been extracted for convenience.

We assume that the pulses are spectrally 
very broad to assure resonance with all the QDs in the ensemble. 
If the durations of the pulses are much shorter than both
$\hbar/\Delta$ and $\hbar/V$, the action of each of them corresponds 
to an instantaneous, independent rotation of the state of each QD,
that is, to the  
unitary transformation $\mathsf{U}_{l}=U_{l}\otimes U_{l}$, where 
\begin{eqnarray*}
\lefteqn{U_{l}=}\\
&&\cos\left(\frac{\alpha_{l}}{2}\right)\mathbb{I}
-i\sin\left(\frac{\alpha_{l}}{2}\right)
\left[ e^{-i(\phi_{l}+Et_{l}/\hbar)}|0\rangle\!\langle 1|
+\mathrm{h.c.}\right].
\end{eqnarray*}
Here $\mathbb{I}$ denotes the identity operator and
\begin{equation*}
\alpha_{l}=\frac{1}{\hbar}\int_{-\infty}^{\infty}f_{l}(t)dt
\end{equation*}
is the pulse area. 

We assume that the initial state of a DQD is 
$\rho(-\tau^{-})=|00\rangle\!\langle00|$ ($t^{\pm}$ denotes
just after or before a time instant $t$). 
The DQD is then excited with the first pulse. Just after this pulse, 
the system state is
\begin{eqnarray}
	\rho(-\tau^{+})=\mathsf{U_{1}}\rho(-\tau^{-})\mathsf{U^{\dagger}_{1}}
\end{eqnarray}
and the four matrix elements
related to optical polarizations have the values
\begin{eqnarray*}
\rho_{01}(-\tau^{-}) & = & \rho_{02}(-\tau^{-})=\frac{i}{2}\sin\alpha_{1}
\cos^{2}\frac{\alpha_{1}}{2}e^{-i\phi_{1}+iE\tau/\hbar},\\
\rho_{13}(-\tau^{-}) & = & \rho_{23}(-\tau^{-})=\frac{i}{2}\sin\alpha_{1}
\sin^{2}\frac{\alpha_{1}}{2}e^{-i\phi_{1}+iE\tau/\hbar}.
\end{eqnarray*} 
From now on, we will only keep the terms which are of the first order
in the first pulse area $\alpha_{1}$. In this approximation, the
biexciton coherences do not contribute at this stage of the evolution.

Between the laser pulses, the evolution of the reduced density matrix 
of the charge subsystem is described by the Lindblad equation of the 
form \cite{sitek07a}
\begin{eqnarray}\label{L}
\dot{\rho}=-\frac{i}{\hbar}[H_{\mathrm{X}},\rho]+\mathcal{L}[\rho],
\end{eqnarray}
with
\begin{eqnarray}
	\mathcal{L}[\rho]=\Gamma \left[ \Sigma_{-}\rho\Sigma_{+}
-\frac{1}{2}\{\Sigma_{+}\Sigma_{-},\rho \}_{+}\right],
\end{eqnarray}
where $\Gamma$ is the spontaneous decay rate for an individual dot and 
$\Sigma_{-}=\Sigma^{\dagger}_{+}=\sigma^{\left(1\right)}_{-}+\sigma^{\left(2\right)}_{-}$. 
This yields a closed system of four equations of motion for the
negative frequency parts of optical polarizations,
\begin{subequations}
\begin{eqnarray}\label{evol-eq}
\dot{\rho}_{01} &= &\left(i\Delta/\hbar-\Gamma/2 \right)\rho_{01}
+\left( iV/\hbar- \Gamma/2\right) \rho_{02}\nonumber \\
&&+\Gamma(\rho_{13}+\rho_{23}), \\
\dot{\rho}_{02} &= &\left(iV/\hbar- \Gamma/2\right)\rho_{01}
- \left(i\Delta/\hbar+\Gamma/2 \right)\rho_{02}\nonumber \\
&&+\Gamma(\rho_{13}+\rho_{23}), \\
\dot{\rho}_{23} &= &
\left(iV_{\mathrm{B}}/\hbar+i\Delta/\hbar-3\Gamma/2 \right)\rho_{23}
\nonumber \\
&&-\left(iV/\hbar+ \Gamma/2\right) \rho_{13},\\
\dot{\rho}_{13} &= &-\left(iV/\hbar+ \Gamma/2\right)\rho_{23}\nonumber \\
&&+ \left(iV_{\mathrm{B}}/\hbar-i\Delta/\hbar-3\Gamma/2 \right)\rho_{13}.
\end{eqnarray}
\end{subequations}
The solution to these equation simplifies if one notes that for
typical double dots the energy mismatch is much larger than the
fundamental line width. Therefore, in the following we assume
$\Gamma\ll \Omega/\hbar$, where $\Omega=(\Delta^{2}+V^{2})^{1/2}$ corresponds to
the half-splitting of the single exciton states. This reflects the
actual experimental situation as the energy mismatch observed in
real samples ranges from a few meV to tens of meV
\cite{gerardot03,gerardot05,borri03} while the recombination rate
$1/\Gamma$ is typically of the order of 1~ns
\cite{borri03,langbein04,gerardot05}. 

Upon solving the equations (\ref{evol-eq}-d), one finds the exciton
coherences at $t=0^{-}$, 
\begin{eqnarray}\label{0-}
\lefteqn{\rho_{01(02)}(0^{-})  = \frac{i}{4}\sin\alpha_{1}\cos^{2}\frac{\alpha_{1}}{2}
e^{-i\phi_{1}+iE\tau/\hbar}}\\
&&\times\left[ \left( 1-\frac{V\pm\Delta}{\Omega} \right)e^{\lambda_{0}\tau}
+\left( 1+\frac{V\pm\Delta}{\Omega} \right)e^{\lambda_{1}\tau}
\right],
\nonumber
\end{eqnarray}
with
\begin{equation}\label{lambda01}
\lambda_{0(1)}=\mp i\frac{\Omega}{\hbar}-\frac{1}{2}\beta_{\mp}\Gamma,
\end{equation}
where $\beta_{\pm}=1\pm V/\Omega$, and the upper sign corresponds to the
first index or pair of indices.

The second pulse, with an area $\alpha_{2}$ arrives at $t=0$. In order
to find the FWM response to the leading (third) order, we need to
calculate the positive frequency parts of optical polarizations, keeping only
terms of the second order, containing the phase factor
$e^{2i\phi_{2}}$. Such terms depend only on the values of the
negative frequency polarizations before the pulse and have the form
\begin{subequations}
\begin{eqnarray}
\rho_{10}(0^{+}) & = & \rho_{20}(0^{+})\nonumber\\
&=&\frac{e^{i\phi_{2}}}{4}\sin^{2}\alpha_{2}
\left[ \rho_{01}(0^{-})+\rho_{02}(0^{-}) \right],\label{0+a}\\
\rho_{31(32)}(0^{+}) & = & -\frac{e^{i\phi_{2}}}{4}\sin^{2}\alpha_{2}
\rho_{02(01)}(0^{-}).\label{0+b}
\end{eqnarray}
\end{subequations}
Note that biexciton polarizations appear in the leading order and
cannot be eliminated since,
contrary to the single dot case, the transition to the molecular biexciton
(excitons with the same polarization confined in
different dots) is not forbidden by selection rules for any
polarization of the laser beam.

The single exciton polarizations at an arbitrary time $t>0$ are found
by using  Eqs.~(\ref{0+a},b) as initial
values for the system of equations of motion for positive frequency
polarizations, which is obtained from Eqs.~(\ref{evol-eq}-d) by
complex conjugation. The total single-exciton polarization is 
$P^{(1)}(t)=\rho_{10}(t)+\rho_{20}(t)$ and is explicitly given by
\begin{eqnarray}
P^{(1)}(t)& = & \frac{i}{8}\sin\alpha_{1}\cos^{2}\frac{\alpha_{1}}{2}
\sin^{2}\alpha_{2}e^{i(2\phi_{2}-\phi_{1})}e^{iE\tau/\hbar}\nonumber\\
&&\times 
\left( \beta_{-}e^{\lambda_{0}^{*}t}+\beta_{+}e^{\lambda_{1}^{*}t}\right)
\left( \beta_{-}e^{\lambda_{0}\tau}+\beta_{+}e^{\lambda_{1}\tau}\right).
\label{P1ex}
\end{eqnarray}
Although, in principle, the biexciton term
affects the exciton coherences in the same order of the optical
response (due to radiative recombination) one finds that the
corresponding terms are of the order of $\hbar\Gamma/\Omega\ll 1$ and can be
neglected. 

For the biexciton polarizations, the evolution equations
(\ref{evol-eq}-d) yield 
\begin{eqnarray*}
\lefteqn{\rho_{31(32)}=} \\
&& \frac{1}{2}\left[
\left( 1\mp\frac{\Delta}{\Omega} \right)\rho_{31(32)}(0^{+})
+\frac{V}{\Omega}\rho_{32(31)}(0^{+}) \right]e^{\lambda_{2}^{*}t}\\
&&+\frac{1}{2}\left[
\left( 1\pm\frac{\Delta}{\Omega} \right)\rho_{31(32)}(0^{+})
-\frac{V}{\Omega}\rho_{32(31)}(0^{+}) \right]e^{\lambda_{3}^{*}t},
\end{eqnarray*}
where 
\begin{equation}\label{lambda23}
\lambda_{2(3)}=\mp i\Omega/\hbar +iV_{\mathrm{B}}/\hbar
-\left( 1+\frac{1}{2}\beta_{\pm}\right)\Gamma .
\end{equation}
The biexciton contribution to the coherent polarization is then 
\begin{eqnarray}
P^{(2)}(t) & = & \rho_{31}+\rho_{32}\nonumber\\
& = & \frac{i}{8}\sin\alpha_{1}\cos^{2}\frac{\alpha_{1}}{2}
\sin^{2}\alpha_{2}e^{i(2\phi_{2}-\phi_{1})}e^{iE\tau/\hbar}\nonumber\\
&&\times \left( \beta_{+}e^{\lambda_{2}^{*}t+\lambda_{1}\tau} 
+\beta_{-}e^{\lambda_{3}^{*}t+\lambda_{0}\tau}\right).
\label{P2ex}
\end{eqnarray}

\subsection{Ensemble response}

The FWM polarization is obtained upon returning to the Schr\"odinger
picture, which amounts to inserting the phase factor $e^{-iEt/\hbar}$
and adding up the contributions from all the dots in the ensemble
according to their statistical distribution given by Eqs.~(\ref{g})
and (\ref{gA}). For
the sake of the further discussion it is convenient to split the exciton
contribution (\ref{P1ex}) into two parts and to write the total
polarization as a sum of three contributions
\begin{eqnarray}
\lefteqn{P_{\mathrm{FWM}}(t)=}\nonumber \\
&&\frac{i}{8}\sin\left(\alpha_{1}\right)
\cos^{2}\left(\frac{\alpha_{1}}{2}\right)
\sin^{2}\left(\alpha_{2}\right)
e^{-\frac{\Gamma}{2}\left(\tau+t\right)}\nonumber \\
&&\times\int d\Delta dEg(E,\Delta)
e^{-iE\left(t-\tau\right)/\hbar}\sum_{n=1}^{3}P_{n}(t,\Delta)+\mathrm{c.c.}
\nonumber\\
&&=\frac{i}{8}\sin\left(\alpha_{1}\right)
\cos^{2}\left(\frac{\alpha_{1}}{2}\right)
\sin^{2}\left(\alpha_{2}\right)\nonumber\\
&&\times e^{-\frac{\sigma_{E}^{2}\left(t-\tau\right)^{2}}{2}-i\overline{E}
\left(t-\tau\right)/\hbar-\frac{\Gamma}{2}\left(t+\tau\right)}\nonumber \\
&&\times\int d\Delta g(\Delta)\sum_{n=1}^{3}P_{n}(t,\Delta)+\mathrm{c.c.},
\label{Pfwm}
\end{eqnarray}
where we inserted the definitions (\ref{lambda01}) and
(\ref{lambda23}) into Eqs.~(\ref{P1ex}) and (\ref{P2ex}) and
defined
\begin{eqnarray*}
\label{P}
\lefteqn{P_{n}(t,\Delta)=P_{n+}(t,\Delta)+P_{n-}(t,\Delta),}\\
\lefteqn{P_{1\pm}(t,\Delta)=
\beta_{\pm}^{2}\exp\left[ 
\mp \frac{V\Gamma}{2\Omega}\left(\tau+t\right)
\pm i\Omega\left(\tau-t\right)/\hbar  \right],}  \\
\lefteqn{P_{2\pm}(t,\Delta)=
\beta_{\pm}\beta_{-}\exp\left[ 
\pm \frac{V\Gamma}{2\Omega}\left(\tau-t\right)
\mp i\Omega\left(\tau+t\right)/\hbar \right],}  \\
\lefteqn{P_{3\pm}(t,\Delta)=}\\
&&-\beta_{\mp}\exp\left[-\left( \Gamma+iV_{B}/\hbar\right) t
\pm \left(\frac{V\Gamma}{2\Omega}- i\Omega/\hbar\right)\left(\tau+t\right)
 \right].  
\end{eqnarray*}
The second form of
Eq.~(\ref{Pfwm}) is obtained by performing the integration over $E$.

Next, one has to calculate the heterodyne signal generated by the FWM
polarization overlapped with the reference signal, according to
Eq.~(\ref{F}). 
The integration over time in Eq.~(\ref{F}) can easily be
performed. Substituting Eq.~(\ref{Pfwm}) into Eq.~(\ref{F}) and
performing some reasonable approximations, as discussed in detail in
the Appendix, one finds 
\begin{displaymath}
F(t_{0},\tau)=\sum_{n=1}^{3}F_{n}(t_{0},\tau),
\end{displaymath}
where
\begin{eqnarray}
F_{n}(t_{0},\tau) & = & 
\frac{i}{8}\sin\left(\alpha_{1}\right)
\cos^{2}\left(\frac{\alpha_{1}}{2}\right)
\sin^{2}\left(\alpha_{2}\right) \nonumber\\
&&\times\int d\Delta g_{\Delta}(\Delta) 
[\mathcal{F}_{n+}(\Delta)+\mathcal{F}_{n-}(\Delta)].
\label{F2}
\end{eqnarray}
Here
\begin{eqnarray}
\mathcal{F}_{n\pm}(\Delta) & = &
e^{i\overline{E}(t_{0}-\tau)/\hbar} 
\int dt P_{n\pm}^{(+)}(t,\Delta)E_{\mathrm{ref}}^{(-)}(t) \nonumber\\
&=& \frac{1}{4\sqrt{1+\tau_{\mathrm{ref}}^{2}\sigma_{E}^{2}/\hbar}}
\exp\left[
-\frac{\sigma_{E}^{2}(t_{0}-\tau)^{2}}{
2(\hbar^{2}+\tau_{\mathrm{ref}}^{2}\sigma_{E}^{2})} \right] \nonumber\\
&&\times \exp\left[\mp i\frac{\Omega (t_{0}-\tau)}{
\hbar(\hbar+\tau_{\mathrm{ref}}^{2}\sigma_{E}^{2}/\hbar)} \right]
\Phi_{n\pm}(\Delta),
\label{P-Delta}
\end{eqnarray}
where
\begin{subequations}
\begin{eqnarray}
\label{Pi}
\Phi_{1\pm}(\Delta)&=& \beta^{2}_{\pm}
\exp\left[-
\frac{\Omega^{2}\tau^{2}_{\mathrm{ref}}}{
2(\hbar^{2}+\tau_{\mathrm{ref}}^{2}\sigma_{E}^{2})} \right] \nonumber\\
&& \times
\exp\left[-\beta_{\pm}\Gamma\tau\right],\\
\Phi_{2\pm}(\Delta) &=& \beta_{+}\beta_{-}
\exp\left[
-\frac{\Omega^{2}\tau^{2}_{\mathrm{ref}}}{
2(\hbar^{2}+\tau_{\mathrm{ref}}^{2}\sigma_{E}^{2})} \right] \nonumber\\
&&\times 
\exp\left[\mp 2i\Omega\tau/\hbar\right]
\exp\left[-\Gamma\tau\right],\label{Pi-b}\\
\Phi_{3\pm}(\Delta) &=& 
-\beta_{\mp}
\exp\left[
-\frac{(\Omega\pm V_{\mathrm{B}})^{2}\tau^{2}_{\mathrm{ref}}}{
2(\hbar^{2}+\tau_{\mathrm{ref}}^{2}\sigma_{E}^{2})} \right] \nonumber\\
&&\times\exp\left[-i(V_{\mathrm{B}}\pm 2\Omega)\tau/\hbar
-\frac{iV_{\mathrm{B}}(t_{0}-\tau)}{(\hbar+\tau_{\mathrm{ref}}^{2}\sigma_{E}^{2}/\hbar)}
\right] \nonumber\\
&&\times 
\exp\left[-(1+\beta_{\mp})\Gamma\tau \right].
\label{Pi-c}
\end{eqnarray}
\end{subequations}

In order to calculate the time-resolved nonlinear response of
the DQD ensemble, Eq.~(\ref{P-Delta}) must be integrated numerically
with the distribution function
$g_{\Delta}(\Delta)$, according to Eq.~(\ref{F2}). Another
integration, over $t_{0}$, yields the 
time-integrated (TI) FWM signal as a function of the delay time
$\tau$, which is commonly used to characterize the dephasing in a
physical system. The results will be discussed in the following
section.

\section{Results}
\label{sec:results}

In this section we present and discuss the time-resolved and
time-integrated FWM response from an ensemble of double quantum dots,
depending on the statistical distribution of the energy mismatch
between the dots in the ensemble and on the strength of the coupling
between the dots. We 
assume fixed values of the average energy mismatch
$\bar{\Delta}=4$~meV, the standard deviation of the mean DQD transition energy
$\sigma_{E}=8$~meV, the length of the reference pulse
$\tau_{\mathrm{ref}}=21$~fs (corresponding to 50~fs full width at half
maximum), and the spontaneous recombination rate for an individual dot
$\Gamma=1$~ns$^{-1}$. We start by studying the nonlinear response for
short (picosecond) delays; then we proceed to the discussion of the
signal decay on long (nanosecond) time scales.

\subsection{FWM response for short delay times}

\begin{figure}[tb]
\includegraphics[width=85 mm]{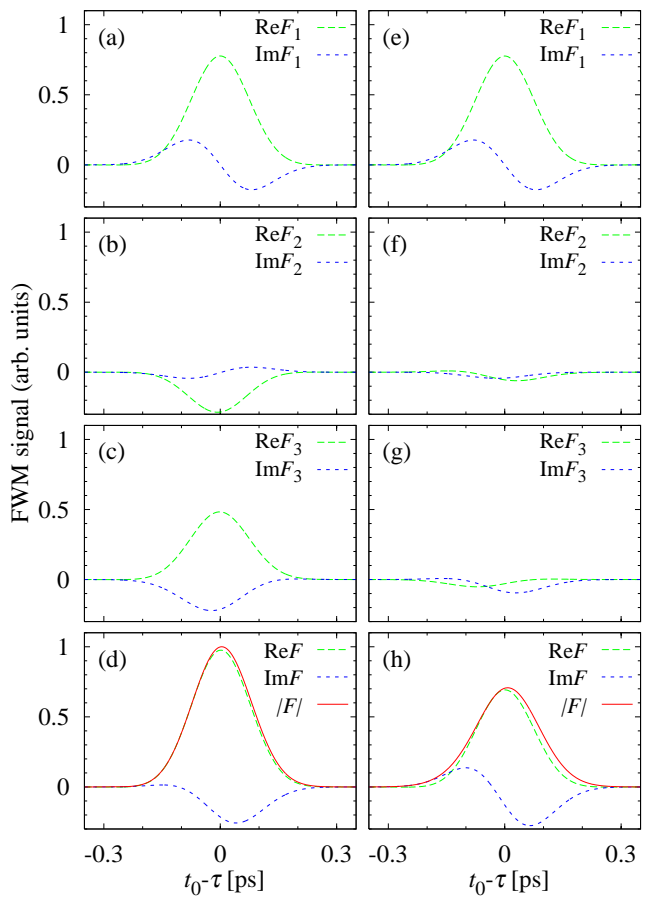}
\caption{\label{fig:echo}(Color online) (a-c) Real and imaginary parts (green dashed
  and blue dotted lines, respectively) of the three
contributions $F_{n}(t_{0},\tau)$ to the FWM echo $F(t_{0},\tau)$  
[Eqs.~(\ref{F2})--(\ref{Pi-c})] for
short delays, with $\tau=0.23$~ps, $V=2$~meV,
$V_{\mathrm{B}}=1$~meV, $\sigma_{\Delta}=1$~meV. (e-g) As previously
but at $\tau=0.61$~ps. (d,h) Real and imaginary parts of the total
signal $F(t_{0},\tau)$, as well as its modulus, which corresponds to the measured
signal, for the two values of $\tau$.}
\end{figure}

\begin{figure}[tb]
\includegraphics[width=85 mm]{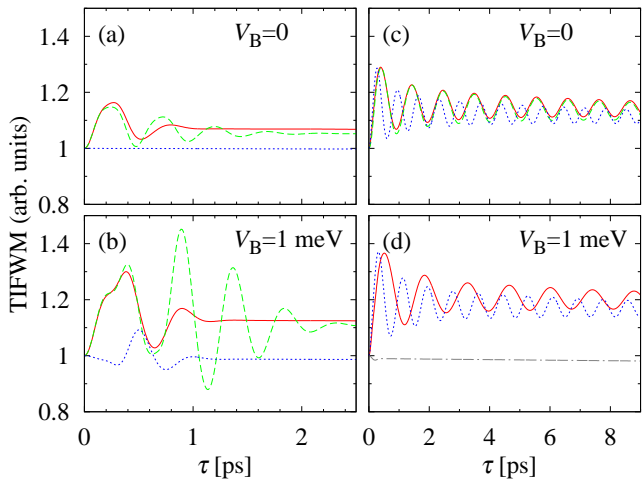}
\caption{\label{fig:Ti}(Color online) (a,b) The evolution of the time-integrated signal for
short delays without (a) and with (b) a biexciton shift for 
$\sigma_{\Delta}\ll\overline{\Delta}$ (no resonant dots in the
ensemble). Red solid 
lines: $V=2$~meV, $\sigma_{\Delta}=1$~meV; green dashed lines:
$V=2$~meV, $\sigma_{\Delta}=0.5$~meV; blue dotted lines: $V=0$~meV,
$\sigma_{\Delta}=1$~meV. (c,d) As in (a) and (b), respectively, but
for $\sigma_{\Delta}$ comparable with $\overline{\Delta}$. Red solid
lines: $V=2$~meV, $\sigma_{\Delta}=3$~meV; green dashed line:
$V=2$~meV, $\sigma_{\Delta}=5$~meV; blue dotted lines: $V=3$~meV,
$\sigma_{\Delta}=3$~meV; grey dash dotted line $V=0$, 
$\sigma_{\Delta}=3$~meV.}
\end{figure}

It follows from Eq.~(\ref{P-Delta}) that all the contributions to the
FWM signal are restricted to the short range of delay times around
$t_{0}=\tau$, of width $\sim\hbar/\sigma_{\mathrm{E}}$, that is, they
have the form of a photon echo. However, 
among the three contributions $F_{n}$ [Eq.~(\ref{F2})], the first one,
$F_{1}$, has a different character than the other two. 
This results from the different structure
of the phase factors of the form $e^{i\Omega\tau/\hbar}$ appearing in
the last exponential term of 
Eq.~(\ref{P-Delta}), in the second exponential term of
Eq.~(\ref{Pi-b}) and in the second exponential term of 
Eq.~(\ref{Pi-c}). Since $\Omega$ depends on $\Delta$, which varies
across the ensemble, such terms
tend to interfere destructively when the signal from different DQDs
is added up. However, in the case of $F_{1}$, this phase term
depends only on $t_{0}-\tau$, which is limited to the width
$\hbar/\sigma_{\mathrm{E}}$ of the echo pulse. Therefore, the spread
of the phase factors is also limited and independent of $\tau$. The
only effect of this phase term is therefore a slight asymmetry of the
echo pulse due to oscillations in $\im F_{1}$, as shown in
Fig.~\ref{fig:echo}(a,e) 
(which makes it different from the simple pulse shape
from an ensemble of individual dots). These oscillations
are always in phase with the center of the echo
peak, so that the peak area (that is, the time-integrated signal) is
constant. 

The situation is different in the case of the other two
contributions $F_{2}$ and $F_{3}$. 
Here, another phase term appears, proportional to
$\Omega\tau/\hbar$ or $(V_{\mathrm{B}}\pm 2\Omega)\tau/\hbar$ 
[in the second exponential terms of Eqs.~(\ref{Pi-b}) and (\ref{Pi-c})]. 
Thus, the phase of the signal at
$t_{0}=\tau$ varies with $\tau$, which leads to a variation of the
shape and magnitude of the echo. As can be seen by comparing 
Figs.~\ref{fig:echo}(b,c)] with Figs.~\ref{fig:echo}(f,g)], 
the contribution of these terms to the photon echo strongly depends on
the delay $\tau$. As a result, also the amplitude of the total
(actually measured) signal varies with time on picosecond time scales,
as shown in Figs.~\ref{fig:echo}(d,h). This variation leads to  
oscillations 
in the TI signal (Fig.~\ref{fig:Ti}). These oscillations are a
manifestation of optical beats between the two dots. Their form 
depends on whether the ensemble contains a
fraction of DQDs composed of identical (resonant) dots, that is, on
the interplay of $\sigma_{\Delta}$ and $\overline{\Delta}$.

If $\sigma_{\Delta}\ll \overline{\Delta}$ then the signal originates
from all the DQDs, whose values of the energy mismatch $\Delta$ lie
roughly within 
$\sigma_{\Delta}$ from $\overline{\Delta}$. The inhomogeneity of the values
of $\Delta$ translates into inhomogeneity of $\Omega$ and leads to a
spread of phases in the terms like $\exp(i\Omega\tau)/\hbar$, which increases
as $\tau$ increases. Since the total signal from the sample is a
coherent sum of the fields emitted by all DQDs, this phase
distribution leads to quenching of the two contributions $F_{2}$
and $F_{3}$ at 
$\tau\sim \hbar\Omega/(\Delta\sigma_{\mathrm{\Delta}})\ll
1/\Gamma$ and, therefore,
to vanishing of the oscillations. This can be clearly seen in
Fig. \ref{fig:Ti}(a,c). This effect is due to the fact that
the probe pulse acts symmetrically on 
both dots and, therefore, can invert (refocus) only the dephasing
due to the inhomogeneous distribution of the average transition
energies but not that resulting from the inhomogeneity of the energy
mismatch between the dots in a single pair. 

The evolution of the optical signal becomes more interesting if
$\sigma_{\Delta}\gtrsim\overline{\Delta}$. This condition means that
the ensemble contains a fraction of resonant DQDs, that is, such that
have nearly identical 
transition energies. The frequency $\Omega$ has a minimum at
$\Delta=0$ which corresponds to a stationary point of the phase
distribution over the ensemble. Therefore, all the nearly resonant QDs
emit radiation in phase and can dominate the contributions $F_{2}$
and $F_{3}$ of the ensemble when the signal from the possibly much
more numerous dots with $\Delta\sim\overline{\Delta}$ has dephased as described
above. An essential point is that, since this minority resonant
subensemble has $\Delta\sim 0$, the frequency of the beats is very close
to $2V/\hbar$ or $(2V-V_{\mathrm{B}})/\hbar$ [note that the term with this frequency
has a much larger amplitude in Eq.~(\ref{Pi-c}) than that with the
frequency $(2V+V_{\mathrm{B}})/\hbar$ since for $\Delta\to 0$ one has
$\beta_{-}\to 0$]. In this way, the nonlinear response 
gives a direct
access to the values of the inter-dot couplings and to the properties of
resonant dots, even if they are a minority in the ensemble.
The beats originating from the resonant subensemble show slower
damping than those from the majority DQDs but still they vanish on
the time scales of several picoseconds.

\subsection{FWM response for long delays}

For longer delays, only $F_{1}$
contributes to the signal. Out of the two components making up this
term, $F_{1+}$ has the decay rate 
$\Gamma_{+}=\beta_{+}\Gamma>\Gamma$. This is the superradiant
component of the optical coherence, the decay rate of which reaches
$2\Gamma$ when $V\gg \Delta$. The other, subradiant component $F_{1-}$
has the 
decay rate $\Gamma_{-}=\beta_{-}\Gamma<\Gamma$, which vanishes in the
limit of strongly coupled dots. The relative amplitude of these two
components is $\beta_{+}/\beta_{-}$, so that the superradiant one
dominates if the coupling is strong, with a subradiant tail visible
only at long delays.

\begin{figure}[tb]
\includegraphics[width=85 mm]{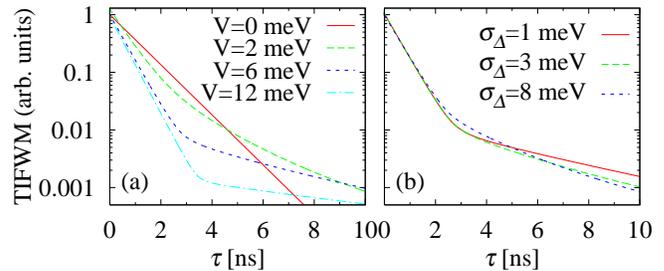}
\caption{\label{fig:dlugie}(Color online) Time-integrated FWM response. (a)
Dependence on the coupling strength $V$ for
$\sigma_{\Delta}=3$~meV. (b) Dependence on the inhomogeneity of the
energy mismatch for $V=5$~meV.}
\end{figure}

The decay of the FWM signal at long delay times
is shown in Fig. \ref{fig:dlugie}(a)
for different strengths of coupling between
the dots. For $V=0$, both decay
rates are equal to $\Gamma$ and one observes a usual exponential
decay. As the coupling increases, the decay
of the FWM response becomes non-exponential due to the presence of the
sub- and superradiant components. 
For $V\sim\Delta$, the decay sets off with an intermediate rate
between $\Gamma$ and $2\Gamma$ but it deviates from an exponential
form rather quickly. When $V\gg
\Delta$ the FWM response is almost completely dominated by the
superradiant component, showing an exponential decay with the rate
$2\Gamma$ over a very long time range, with only a weak tail corresponding to the
subradiant part of the signal. 

Since the decay rates $\Gamma_{\pm}$ depend on $V/\Omega=V/(V^{2}+\Delta^{2})^{1/2}$, which
varies over the ensemble due to the inhomogeneity of $\Delta$, they
change from one DQD to another. In order to see how this
inhomogeneity effect influences the nonlinear optical response we
plot the time-integrated FWM response for a few values of
$\sigma_{\Delta}$ in Fig.~\ref{fig:dlugie}(b). The influence of this
parameter is rather small and appears only for rather long times. The
major effect is some softening of the shoulder which marks the transition
from the dominating superradiant contribution to the subradiant
tail. However, this happens only when $\sigma_{\Delta}\gg\Omega$.

Interestingly, as long as only the leading order response is
considered, the biexciton shift appears only in the short-living term
$F_{3}$. At longer times, this kind
of coupling does not affect the FWM response at all. Thus, there are
no beats from molecular biexcitons in the FWM response.

\section{Conclusion}
\label{sec:concl}

Our results show that the shape and decay rate of the time-resolved 
FWM signal provide rich information on the properties of the DQDs
in the ensemble, including coupling between the
QDs. In the time-integrated signal for long delay times, 
one observes a transition from the
regime of independent decay (with the usual decay rate) to superradiant 
decay (double rate) via intermediate cases of non-exponential 
decay. This transition is driven by the interplay of 
the energy mismatch between the QDs forming the DQD system and the coupling 
between them and is only weakly affected by the distribution of the
energy mismatch. The superradiance effect observed in the decay of the
FWM response is strongly stabilized by the coupling between the dots
and can survive even when the energy mismatch is many orders of
magnitude larger than the radiative line width.

Unlike ensembles of individual QDs, the DQD samples show interesting
and meaningful features also in the TI signal at short delays (several
picoseconds). In this range of delay times, the signal has the form of
oscillations which result from optical beats between the radiation
emitted from different DQDs.  The form and decay time of these
oscillations depends on the presence of a resonant
subensemble (a subset of DQDs with matched transition energies) in the
inhomogeneous ensemble. If such subensemble is present, its
contribution dominates the short-delay response yielding a direct access to the
strength of interaction between the dots forming the DQDs. 

In our study, we assumed that the surface density of DQDs in the ensemble is not
very high and the distribution of the transition energies is rather
broad so that collective effects on the level of the ensemble are
absent. Including such effects would require applying a more general
theory  \cite{milonni74,milonni75,slepyan01}, as
the DQDs are spaced by distances comparable or larger than the emitted
wave length so that propagation and retardation effects would come into
play.

\begin{acknowledgments}
This work was supported by the Polish MNiSW under Grant No. N~N202~1336~33. 
\end{acknowledgments}

\appendix

\section{Third order response: approximations}

In this Appendix we give the full formulas for the components of the
third order response and present the details of the approximations
that lead to Eqs.~(\ref{Pi}-c). Substituting Eq.~(\ref{Pfwm}) into
Eq.~(\ref{F}) one finds the FWM response signal in the form of
Eqs.~(\ref{F2}) and (\ref{P-Delta}), with
\begin{subequations}
\begin{eqnarray}
\label{Pi-complete}
\Phi_{1\pm}(\Delta)&=& 
\beta^{2}_{\pm}
\exp\left[
\frac{(2i\Omega/\hbar\pm\beta_{\pm}\Gamma)^{2}\tau^{2}_{\mathrm{ref}}}{
8(1+\tau_{\mathrm{ref}}^{2}\sigma_{E}^{2}/\hbar^{2})} \right] \nonumber\\
&& \times
\exp\left[-\beta_{\pm}\Gamma\tau
-\frac{\beta_{\pm}\Gamma(t_{0}-\tau)}{
2(1+\tau_{\mathrm{ref}}^{2}\sigma_{E}^{2}/\hbar^{2})} \right],
\\
\Phi_{2\pm}(\Delta) &=& 
\beta_{+}\beta_{-}
\exp\left[
\frac{(2i\Omega/\hbar\pm\beta_{\pm}\Gamma)^{2}\tau^{2}_{\mathrm{ref}}}{
8(1+\tau_{\mathrm{ref}}^{2}\sigma_{E}^{2}/\hbar^{2})} \right] \nonumber\\
&& \times
\exp\left[\mp 2i\Omega/\hbar\tau\right] \nonumber \\
&&\times 
\exp\left[-\Gamma\tau
-\frac{\beta_{\pm}\Gamma (t_{0}-\tau)}{
2(1+\tau_{\mathrm{ref}}^{2}\sigma_{E}^{2}/\hbar^{2})} \right],
\\
\Phi_{3\pm}(\Delta) &=& 
-\beta_{\mp}\exp\left[
\frac{(2i\Omega/\hbar\pm\left(2+\beta_{\mp}\right)\Gamma)^{2}\tau^{2}_{\mathrm{ref}}}{
8(1+\tau_{\mathrm{ref}}^{2}\sigma_{E}^{2}/\hbar^{2})} \right] \nonumber\\
&&\times 
\exp\left[
-\frac{\tau_{\mathrm{ref}}^{2}V_{B}(V_{B}\pm 2\Omega)}{
2(\hbar^{2}+\tau_{\mathrm{ref}}^{2}\sigma_{E}^{2})} \right] \nonumber\\
&&\times \exp\left[
i\frac{(2+\beta_{\mp})\Gamma\tau_{\mathrm{ref}}^{2}V_{B}}{
2(\hbar+\tau_{\mathrm{ref}}^{2}\sigma_{E}^{2}/\hbar)} \right] \nonumber\\
&&\times\exp\left[-i(V_{\mathrm{B}}\pm 2\Omega)\tau/\hbar
-\frac{iV_{\mathrm{B}}(t_{0}-\tau)}{\hbar+\tau_{\mathrm{ref}}^{2}\sigma_{E}^{2}/\hbar}
\right] \nonumber\\
&&\times 
\exp\left[-(1+\beta_{\mp})\Gamma\tau
-\frac{(2+\beta_{\mp})\Gamma(t_{0}-\tau)}{
2(1+\tau_{\mathrm{ref}}^{2}\sigma_{E}^{2}/\hbar^{2})} \right].
\label{Pi-c-complete}
\end{eqnarray}
\end{subequations}

The values of the radiative decay rate $\Gamma$ and
the inhomogeneous ensemble broadening of the transition energies
$\sigma_{\mathrm{E}}$ are of the order of $\mu$eV and tens of meVs,
respectively, while the typical duration of a reference pulse is about
100~fs or less. Based on these values, Eqs.~(\ref{Pi-complete}-c) can
be considerably simplified.

In the first exponents in Eqs.~(\ref{Pi-complete},b), one can
write (consistently with the approximation $\Gamma\ll\Omega/\hbar$ used
throughout this paper)
\begin{equation*}
(2i\Omega/\hbar\pm\beta_{\pm}\Gamma)^{2}\tau^{2}_{\mathrm{ref}}
\approx -4\Omega^{2}\tau^{2}/\hbar^{2}\pm4i\beta_{\pm}\Gamma\tau^{2}_{\mathrm{ref}}.
\end{equation*}
Moreover, typically $\Gamma\tau_{\mathrm{ref}}\sim 10^{-4}$ and 
$\Omega\tau_{\mathrm{ref}}/\hbar\sim 0.1$, hence the imaginary part can be
safely neglected. The same argument holds for the first exponential
term in Eq.~(\ref{Pi-c-complete}), so that in all three exponents this
term can be replaced by 
$\exp[-(1/2)\Omega^{2}/(\hbar^{2}+\tau_{\mathrm{ref}}^{2}\sigma_{E}^{2})]$.

Because of the Gaussian term in Eq.~(\ref{P-Delta}), it is clear that
the measured signal is of considerable magnitude only when 
$|t_{0}-\tau|\lesssim \hbar/\sigma_{E}$,
which reflects the `photon echo' nature of the FWM
response.  Since $\hbar\Gamma/\sigma_{\mathrm{E}}\ll 1$,
the very last terms in Eqs. (\ref{Pi-complete}-c),
proportional to $\Gamma(t_{0}-\tau)$, can be discarded. 
The third exponential term in Eq.~(\ref{Pi-c-complete}) is also
negligible, as $\tau_{\mathrm{ref}}\Gamma\ll 1$, while
$\tau_{\mathrm{ref}}V_{\mathrm{B}}/\hbar$ is typically of the order of
0.1 (for a biexciton shift of a few meV). With these approximations
one arrives at the Eqs. (\ref{Pi}-c).


\end{document}